\documentclass[pdflatex,sn-mathphys]{sn-jnl} 
\jyear{2021}
\theoremstyle{thmstyleone}

\theoremstyle{thmstyletwo}%

\theoremstyle{thmstylethree}%

\begin{document}

\title[Fully suspended nano-beams]{Fully suspended nano-beams for 
quantum fluids
%
}

\author[1]{\fnm{Ilya} \sur{Golokolenov}}

\author[1]{\fnm{Baptiste} \sur{Alperin}}
\author[1]{\fnm{Bruno} \sur{Fernandez}}
\author[1]{\fnm{Andrew} \sur{Fefferman}}
\author*[1]{\fnm{Eddy} \sur{Collin}}\email{eddy.collin@neel.cnrs.fr}

\affil*[1]{\orgdiv{Univ. Grenoble Alpes}, \orgname{Institut N\'eel}, \orgaddress{\street{25 rue des Martyrs}, \city{Grenoble}, \postcode{38042}, \country{France}}}

\abstract{ Non-invasive probes are keystones of fundamental research. 
Their size, and maneuverability (in terms of e.g. speed, dissipated power) define their applicability range for a specific use. 
As such, solid state physics possesses e.g. Atomic Force Microscopy (AFM),  Scanning Tunneling Microscopy (STM), or Scanning SQUID Microscopy. In comparison, quantum fluids (superfluid $^{3}\mathrm{He}$, $^{4}\mathrm{He}$) are still lacking probes able to sense them (in a fully controllable manner) down to their smallest relevant lengthscales, namely the coherence length $\xi_0$.
In this work we report on the fabrication and cryogenic 
characterization of fully suspended (hanging over an open window, with no substrate underneath) $Si_3 N_4$ nano-beams, 
of width down to $50 \ \mathrm{nm}$ and quality factor up to $10^5$. 
As a benchmark experiment we used them to investigate the Knudsen boundary layer of a rarefied gas: $^{4}\mathrm{He}$ at very low pressures.
The absence of the rarefaction effect due to the nearby chip surface discussed in Gazizulin et al. \cite{Rasul} is attested, while we report on the effect of the probe size itself. 
}


\maketitle
\newpage
\section{Introduction}\label{sec1}

Quantum fluids, such as $^4 \mathrm{He}$ and $^3 \mathrm{He}$, are unique model systems
tackling complex fundamental questions of many-body physics, from quantum turbulence \cite{turbul} to topological superfluidity (with the detection of elusive Majorana fermions or other exotic quasi-particles) \cite{majorana}, with applications to other fields of fundamental research like dark matter detection \cite{darkmatter1}. 
Many emergent phenomena can be studied in almost ideal conditions, see e.g. Ref. \cite{droplet}.
However, physical outcomes of experiments are limited by the quality of the probes in use, which rely on many drastic requirements concerning sensitivity, non-invasivity, or response-time among the most obvious ones.

Currently, a broad variety of devices has already been implemented to probe quantum liquids down to ultra-low temperatures. Restricting the discussion to mechanical objects immersed in the fluids, one can cite: microspheres \cite{spheres}, vibrating wires \cite{wires,Bradley}, quartz tuning forks \cite{quartzforkfirst,quartzforkdavid, andyforks}, microelectromechanical (MEMS) \cite{Defoort} and nanoelectromechanical (NEMS) \cite{andy1,andy2,andy3,zmeev} systems. All of these tools have their advantages and disadvantages. 
The beam-like NEMS realised today in the clean room can be considered to be the most convenient and the least invasive probes, in particular due to their diameter, which is orders of magnitude smaller than all other mentioned options ($240 \ \mathrm{\mu m}$ for microspheres, $5 \ \mathrm{\mu m}$ for NbTi vibrating wires, etc.). 
The mechanical resonances reach 
high quality factors, with very small masses, a very broad range of working frequencies and high stability.
Besides, these systems are used not only for probing quantum liquids, but also in a variety of growing fields like microwave optomechanics (and quantum electronics) \cite{RegalNatPhys2008}, single-particle mass spectrometry \cite{massspectr} and others. 

When making a suspended beam NEMS device on a chip, there is in practice a finite distance to the underlying substrate \cite{andy1,andy2,andy3,zmeev}. This means that the moving sensor can couple to the chip surface through the fluid it is probing: a finite size effect that can be relevant (or even dominant), depending on the fluid characteristics. 
Such a phenomenon had been demonstrated with $^{4}\mathrm{He}$ gas at $4 \ \mathrm{K}$ in the very low pressure limit in recent papers \cite{Martial,Rasul}. 
What is reported is a decrease of the gas damping onto the beam NEMS, which is understood in terms of a rarefaction phenomenon occurring next to a wall, within a distance of the order of the particle's mean-free-path: the so-called boundary (or Knudsen) layer.
If one wishes to completely avoid this effect, it requires creating and using 
fully suspended systems, where the distance to any surface can be effectively taken as infinite. This is our Motivation I for our new type of probes.

Besides, to avoid as much as possible any disturbance of the fluid by the probe itself, there is a need to decrease the characteristic size of it down to the smallest relevant lengthscale characterizing the quantum liquid: for superfluids this shall be the coherence length $\xi_0$. 
For $^4 \mathrm{He}$ this is a very strong constraint, since $\xi_0$ is on the scale of an atom.
However for $^3 \mathrm{He}$, it varies from about $15 \ \mathrm{nm}$ to $80 \ \mathrm{nm}$ depending on pressure \cite{Vollhardt}, which is of the order of the smallest NEMS beams diameter achievable by modern nanofabrication technologies. This is our Motivation II.

Here we report on the design, fabrication, cryogenic characterisation and preliminary application in $^{4}\mathrm{He}$ gas of fully suspended nano-beams with a size down to $50 \ \mathrm{nm}$ and quality factors up to $ 10^5$, for mechanical resonances in the $\mathrm{MHz}$ range.
The particularity of the design lies also in multiplexing: by connecting the NEMS beams in series, we are able to measure up to 5 resonators in a single cool-down. This ability is obviously scalable to much larger numbers, and is a mandatory capability in some of the modern quantum fluid research \cite{vik}.
Our results on the rarefied gas demonstrate the absence of the boundary layer effect reported in Ref. \cite{Rasul} for fully suspended beams, as it should. 
However, we show that the size of the probe itself leads to a deviation from the standard expected molecular gas damping, which we interpret as a boundary layer phenomenon occurring at the device.

\section{Results}\label{sec2}
\subsection{Design and fabrication}

\begin{figure}[H]
\centering
\includegraphics[]{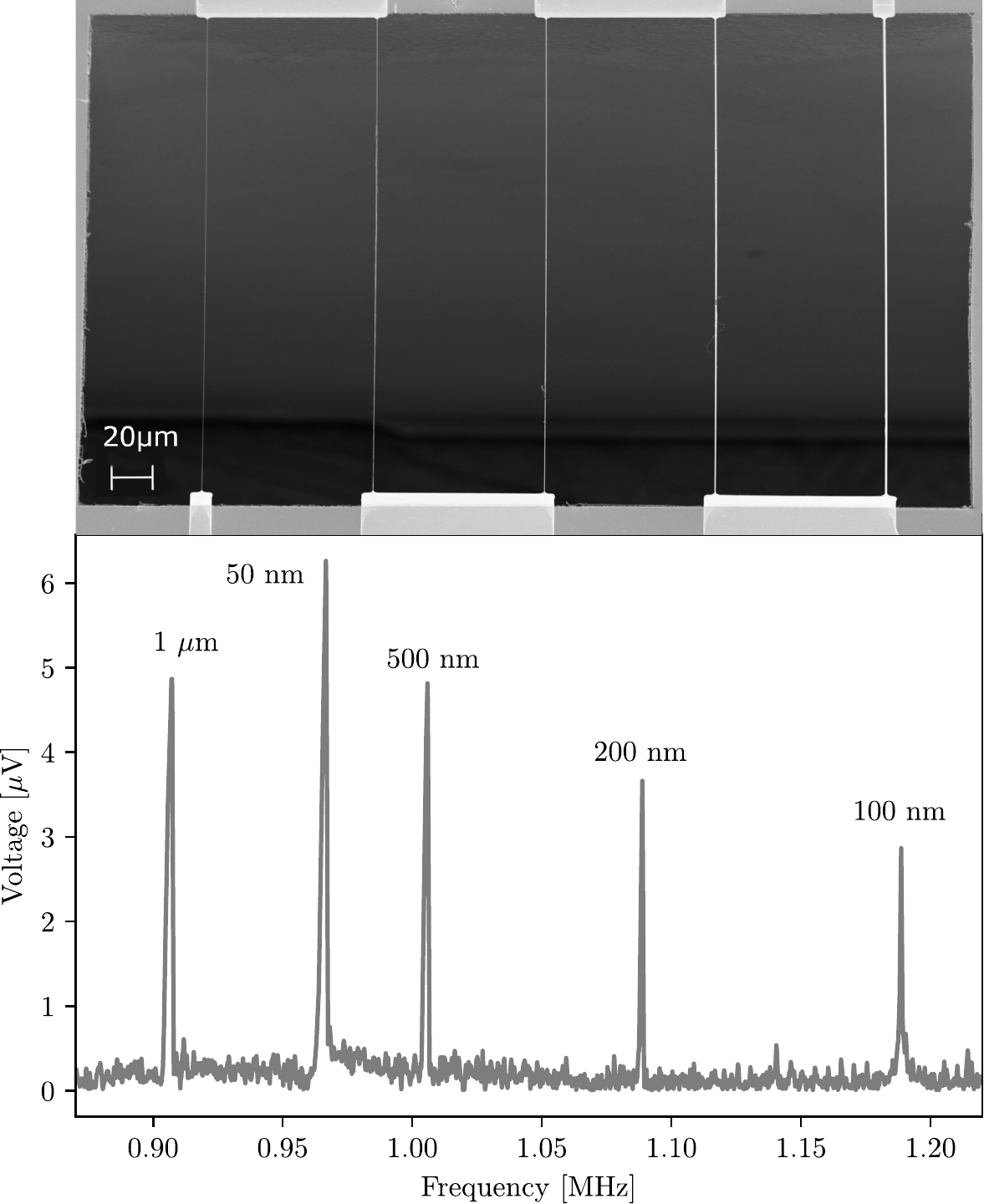}
\caption{{\bf SEM picture of typical sample and all resonances measured in one upward scan}
\newline
{\bf Top:} tilted Scaning Electron Microscopy (SEM) picture of a sample, similar to the one measured. One sees five $Si_3 N_4$ beams covered with $30  \ \mathrm{nm}$ of $Al$ hanging over a $400 \times 400 \ \mathrm{\mu m^2 }$ square hollow window with widths $w = 50 \ \mathrm{nm}$, $100 \ \mathrm{nm}$, $200 \ \mathrm{nm}$, $500 \ \mathrm{nm}$ and $1 \ \mathrm{\mu m}$ respectively from left to right. Picture was taken at an angle of about $30^{\circ}$ from the plane of the chip to show the border between side wall of the etched window and ground (lighter zone above the $20 \ \mathrm{\mu m}$ scale bar, and darker around it, respectively).
{\bf Bottom:} 4.2 K vacuum measured signal from $100 \ \mathrm{\mu m}$ long beams (connected in series, see text), similar to the $400 \ \mathrm{\mu m}$ ones shown on SEM picture (with beam length and distance between them 4 times smaller). 
Signal was acquired with magneto-motive scheme at a magnetic field of about $1.8 \ \mathrm{T}$ (and drive current of $60 \ \mathrm{nA}$, see text).
All beams were also connected independently to define which beam corresponds to which peak (widths are labeled on the figure). Their frequencies and amplitudes change monotonically together with their width, excluding the $50 \ \mathrm{nm}$ beam, which behaves differently. We tend to link it to the fact that at that scale, the size of aluminum grains ($\approx 20 \ \mathrm{nm}$) becomes comparable to the width, which leads to irreproducibility in fabrication quality.
} 
\label{fig1}
\end{figure}

Fabrication starts with a commercial double side polished $Si$ wafer, covered on both sides with $100 \ \mathrm{nm}$ thick $Si_3 N_4$ with a low-stress of about $150 \ \mathrm{MPa}$. Side 1 is coated with resist for protection of the surface, while side 2 is coated with S1805 resist for laser lithography, patterning a square window. 
Silicon nitride is removed there by means of a $SF_6$ RIE plasma-etch.
This window in $Si_3 N_4$ is then used to wet etch $Si$ in a $KOH$ solution to fabricate a $Si_3 N_4$ membrane on side 1. 
With the help of another similar laser lithography step, we align to the membrane edges and make marks (this time on side 1), which are used in the final e-beam lithography (standard PMMA 3\% resist), where the pattern of the main structure is made. 
Aluminum is deposited in this pattern with an e-beam evaporator (about $30 \ \mathrm{nm}$), and used as a mask for $Si_3 N_4$ in the final step of $SF_6$ RIE plasma etching.
This removes the membrane and leaves our beams hanging over a completely hollow opening in the $Si$ wafer (Fig. \ref{fig1}, top). 
The layered structure of our NEMS beams is thus a bottom layer of $100 \ \mathrm{nm}$ of $Si_3 N_4$ and a top one of $30 \ \mathrm{nm}$ of $Al$ (total thickness $e = 130 \ \mathrm{nm}$). 
Here we report on measurements performed with $L = 100 \ \mu m$ long beams, while in practice we made samples of various beam lengths (but same thicknesses and widths), e.g. $200,\ 300\ \mathrm{and}\ 400 \ \mathrm{\mu m}$, see SEM picture in Fig. \ref{fig1} which demonstrates the longest devices we were able to fabricate.

The pads for electrical connection to the beams were made in such a way that one could measure each of them independently (bonding to individual NEMS), but also all of them together, since they are connected in series within the design. 
That allowed us to measure all NEMS resonances together, obtaining a comb in the output signal (detection scheme described below). Measuring each beam independently, we could also
identify with no ambiguity which resonance peak corresponded to which beam
 (see Fig.\ref{fig1}, bottom).

\subsection{Characterisation}

Device characterisation was carried out in a standard $4 \ \mathrm{K}$ $^4\mathrm{He}$ cryostat equipped with a $2 \ \mathrm{T}$ superconducting magnet.
The setup comprises r.f. lines for both drive and detection of the NEMS, plus a thermometer carefully calibrated in the range $1.5 \ \mathrm{K}$ to $30 \ \mathrm{K}$. 
A Baratron$\textsuperscript{\textregistered}$ pressure gauge enables to measure pressure at room temperature from $10^{-3}$ to $100 \ \mathrm{Torr}$. The connecting pipes to the cell have a diameter large enough to ensure that thermomolecular corrections between $4 \ K$ and $300 \ \mathrm{K}$ are negligible. Characterisations were performed in vaccum ($P < 10^{-6} \ \mathrm{Torr}$ typically), prior to gas introduction (see following section).

\begin{figure}[h]
\centering
\includegraphics[width=\linewidth]{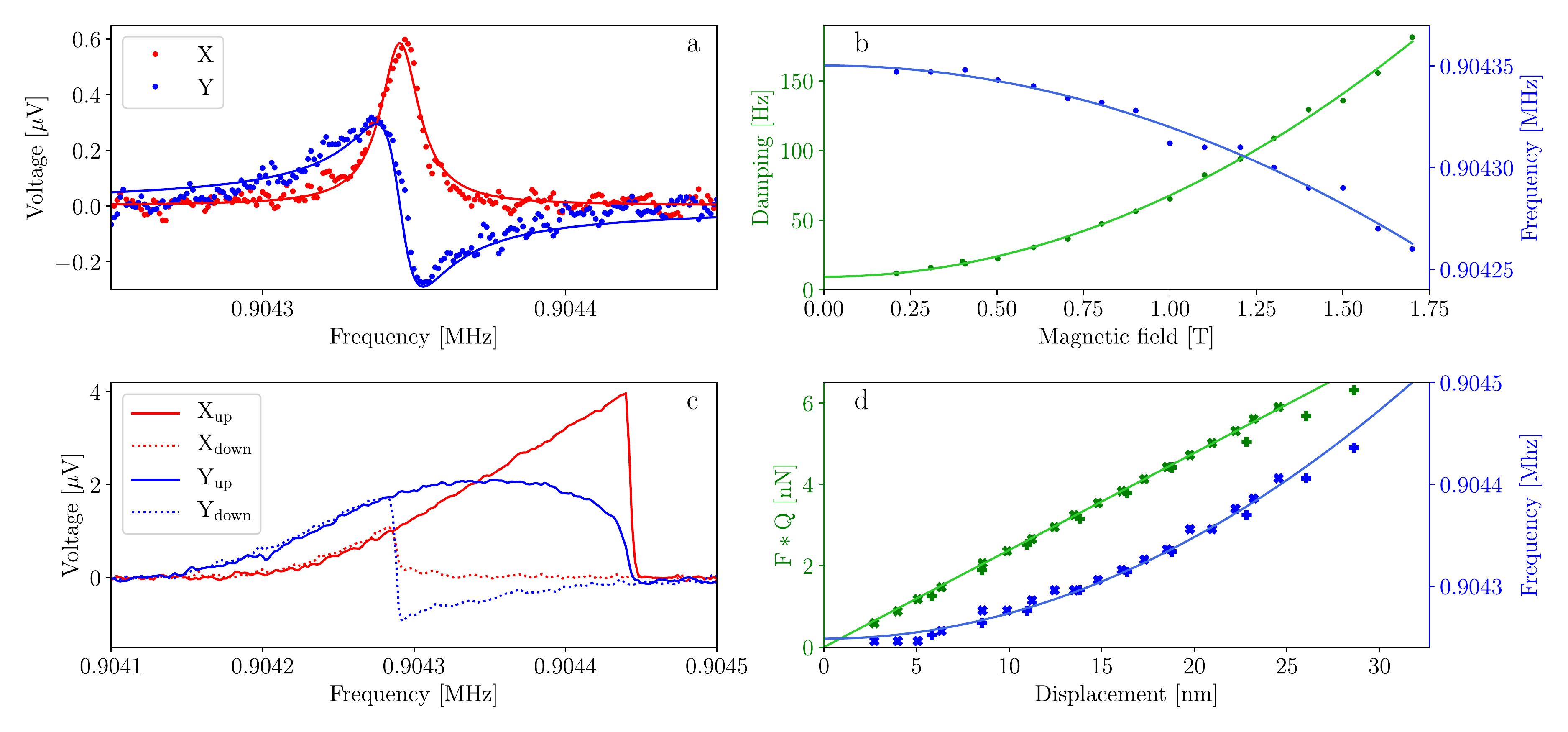}
\caption{{\bf Thorough characterisation of $1 \ \mathrm{\mu m}$ wide, $100 \ \mathrm{\mu m}$ long beam in vacuum at 4.2 K} 
\newline
{\bf a:} Typical linear 
resonance curve, measured at relatively small field and drive with subtracted background ($0.2 \ \mathrm{T}$, $0.5 \ \mathrm{nA}$).
Points correspond to data, while lines demonstrate a Lorentzian fit, from which damping (width of the peak), displacement (height of the peak) and resonance frequency are extracted. 
{\bf b:} Extracted linewidth and resonance frequency, measured at different magnetic fields with fixed drive current (small enough to avoid Duffing nonlinearity at any field). Both are fit with simple quadratic laws (solid lines), which enable to extract intrinsic values: damping $\gamma_0 = 9.3 \ \mathrm{Hz}$ and frequency $f_0 = 904 350 \ \mathrm{Hz}$ (and from the curvature the circuit loading parameters, see text).
This device resonance reaches a quality factor of $Q = 10^5$.
{\bf c:} Typical strongly non-linear curve, measured at relatively small field ($0.4 \ \mathrm{T}$) and high drive ($2.5 \ \mathrm{nA}$) 
displaying hysteresis: increasing scanning frequency (``up") and decreasing it (``down"). 
It is impossible to define the damping from the width of the peak; but we can infer easily the maximal displacement and highest frequency (on  $X_{up}$). 
{\bf d:} Combined plot of force-displacement and frequency-displacement dependencies, measured at two different fields ( $0.4$ and $1.6 \ \mathrm{T}$). 
These dependencies allow us to define spring constant $k = 0.65 \ \mathrm{[N/m]}$ and Duffing parameter $D = 2.5 \cdot 10^{17} \ \mathrm{[Hz/m^2]}$ from the linear and quadratic fits, respectively (see text).
}
\label{fig2}
\end{figure}

Resonator response curves were measured using the well-known magnetomotive technique \cite{Roukes}. 
An a.c. current $I$ is fed in the metallised beam (of length $L$), 
which because of the d.c. field $B$ is subject to a Laplace force $F \propto I \,L \, B$. The motion is detected through the induced voltage $V \propto L \, B \, \dot{x}$, where $\dot{x}$ is the velocity of the beam at its maximal deflection point (out-of-plane motion). The resonance peak associated with the mode is obtained by sweeping the frequency of the drive signal $I$ and detecting the voltage $V$ on a lock-in detector at the same frequency.
In this section we demonstrate typical results for  the beam which has the highest quality factor $Q$. All other beams demonstrate similar behavior. 

The main characterisation results are presented in Fig. \ref{fig2}. In sub-panel {\bf a}, we demonstrate a typical linear response of the beam with its Lorentzian fit.
$\mathrm{X}$ corresponds to the in-phase signal, and $\mathrm{Y}$ to the quadrature signal. From the fit, one can extract the parameters characteristic of linear response, namely damping  $\gamma$ (half-height width of the $\mathrm{X}$ peak), displacement maximum $X_m$ (height of the $\mathrm{X}$ peak) and resonance frequency $f_r$ (position of the maximum on $\mathrm{X}$).

In sub-panel {\bf b} we show the damping and resonance frequency dependence on magnetic field. Measurements were performed at constant drive current, small enough to give a linear response at any applied magnetic field. 
On top of the measured points one can see quadratic fits: $\gamma = \gamma_0 + \beta B^2$ (with $\beta>0$) and $f_r = f_0 + \beta_i B^2$. 
Here, $\gamma_0$ and $f_0$ are intrinsic parameters of the device.
The so-called loading coefficients $\beta$ and $\beta_i$ arise from the 
external circuit, and depend on the actual impedance $R_{ex}+ i\, Z_{ex}$ that the NEMS element sees \cite{Roukes}:
\begin{eqnarray}
\beta \, B^2   & = & f_0 \frac{R_{ex}}{R_{ex}^2+Z_{ex}^2} Z_c, \\
\beta_i \, B^2 & = & \frac{1}{2} f_0 \frac{Z_{ex}}{R_{ex}^2+Z_{ex}^2} Z_c , 
\end{eqnarray}
with $Z_c \propto L^2 B^2 /(m \, 2 \pi f_0)$ the characteristic impedance associated with the beam (field-dependent; $m$ is the mass associated with the mode, see below).
From those fits we find $R_{ex} \approx 280 \ \Omega$ (the setup is high-impedance), 
and $Z_{ex}<0$ (the circuit is capacitive here; the frequency shift is usually not discussed in the literature). The r.f. setup is on purpose not matched to $50 \ \Omega$, in order to minimize the loading effect on $\gamma$.
The intrinsic quality factor is defined as: $Q = f_0/\gamma_0 = 10^5$, which is reasonably high. 

Sub-panel {\bf c} shows a strongly non-linear resonance response, measured in two ways: scanning with increasing frequency (``up"), and decreasing it (``down").
From those resonance curves, we take the frequency of the maximum value (still defined as $f_r$) and its magnitude (again defined as $X_m$) on the $X_{up}$ sweep.
Note the hysteresis between ``up" and ``down", characteristic of the Duffing effect. It arises from the stretching of the beam, as the amplitude of motion grows \cite{LifshitzCross}. 
%

On sub-panel {\bf d} one can see a combined display of the force$\times Q$ vs. displacement relation, and of the Duffing frequency shift due to displacement. These plots include two sets of measurements at two different magnetic fields: $0.4$ and $1.6 \ \mathrm{T}$. 
Quality factors and frequencies for these two fields were taken from sub-panel {\bf b}; in this specific plot, all data overlaps.
$F \, Q$ can be fit linearly, according to:
\begin{equation}
X_m = \frac{F \, Q}{k} ,
\end{equation}
and the slope of the straight line 
directly gives us the spring constant $k$ of the mode. The mode mass is then simply obtained with $m = k/ (2 \pi f_0)^2$.
In a similar way, the Duffing frequency shift can be fit to a parabola \cite{LifshitzCross}:
\begin{equation}
 f_r = f_0 + D \, X_m^2   . 
\end{equation}
We then obtain the Duffing parameter $D$. 
Both numbers are matched to theory (see caption Fig. \ref{fig2} for actual values), leading to a proper calibration of the r.f. lines \cite{MartialPhD}.

\subsection{Gas damping}
The previous section demonstrated excellent mechanical properties of our new devices. In the final part of this paper, we report on their interaction with a benchmark fluid: gaseous $^4$He at $4.2 \ \mathrm{K}$. 
Indeed, $^4$He is a particularly simple gas. It is monoatomic, inert, and can be considered ideal if not too close to the condensation line in the $(P,T)$ phase diagram. Its thermodynamic properties are well known and tabulated (see \cite{Rasul,Martial} and references therein).

\begin{figure}[h]
\centering
\includegraphics[width=\linewidth]{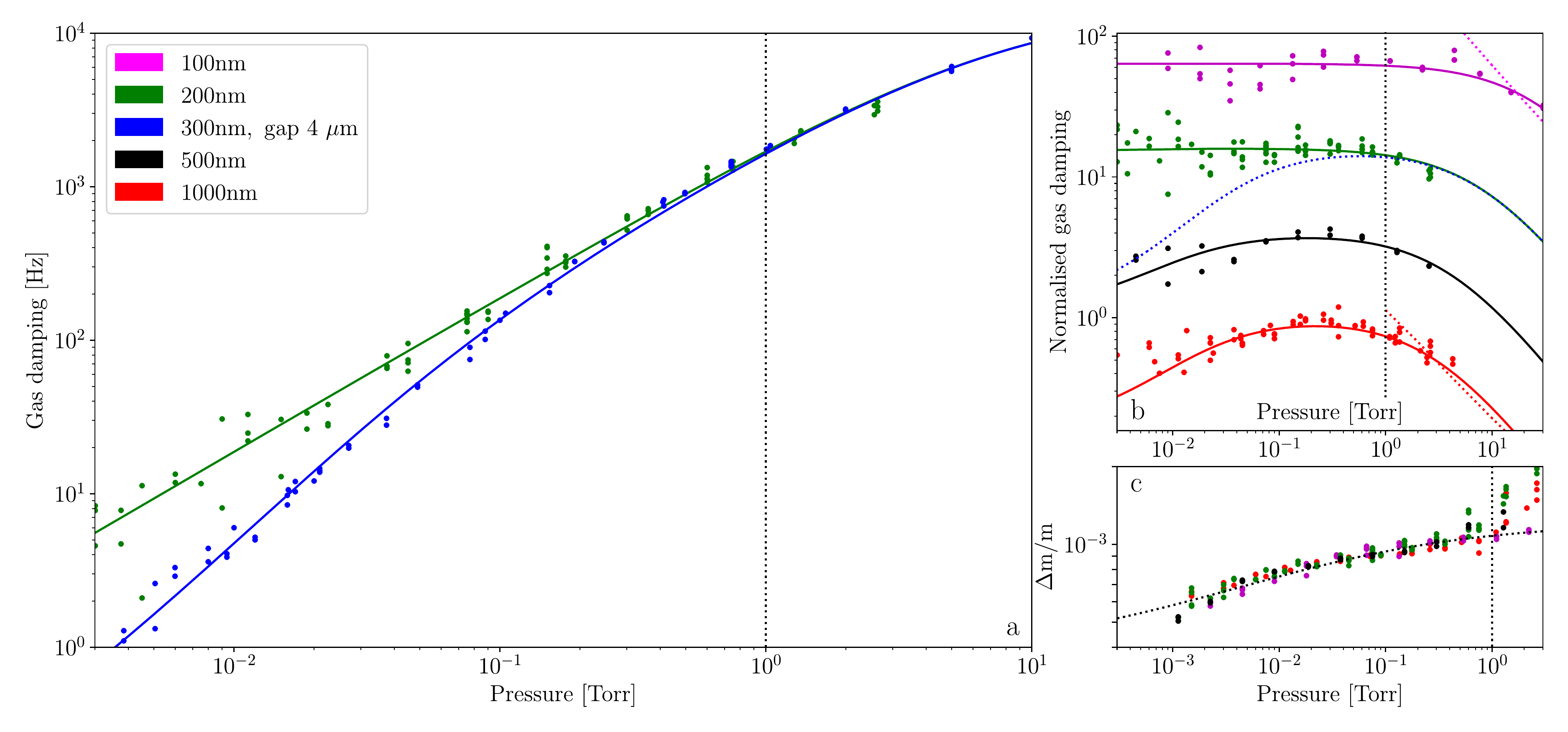}
\caption{{\bf Gas damping at 4.2 K}
\newline
{\bf a:} Comparison of gas damping $\Delta \gamma$ on a fully suspended $200 \ \mathrm{nm}$ wide beam with data from Ref \cite{Rasul} on a $300 \ \mathrm{nm}$ wide device, where the gap to the substrate was about $4 \ \mathrm{\mu m}$. 
{\bf b:} Normalised gas damping measured on beams with different widths: damping was divided by pressure and a constant, chosen such that it equals 1 around a Torr (see text). 
Curves for different devices have then been manually shifted by factors of 4 for clarity. 
Dotted blue line corresponds to fit of the data from Ref \cite{Rasul} (and full line in panel a). All curves were fitted in the same manner (see text; green line same fit in a and b panels). Dotted red and magenta lines indicate theoretical asymptotic tendencies at high pressure  for beams with aspect-ratio width over thickness equal to 10 and 1 respectively (from Ref. \cite{Sader}, see text).
{\bf c:} Frequency shift due to adsorbed atoms on the NEMS surface, re-normalised to relative added mass (see text). Dotted line on top of the data is just a guide for the eyes to underline that all samples behave identically within error-bars.
Vertical dotted lines in all plots indicate pressure equal to $1 \ \mathrm{Torr}$, at which gas damping gradually switches from molecular to laminar regimes.
Note that the legend is common for all subpanels.
}
\label{fig3}
\end{figure}

At high pressures $P$, the fluid is described by the Navier-Stokes theory. The resulting friction can be computed for a rectangular cross section object \cite{Sader}. At low pressures, the fluid is in the molecular regime: the friction is described by individual collisions of atoms with the probe \cite{Bhiladvala}. The cross-over happens for our devices when the mean-free-path of gas particles $\lambda_{mfp}$ equals the width $w$, which is around $1 \ \mathrm{Torr}$ for $w \approx 250 \ \mathrm{nm}$ \cite{Rasul,Martial} (see also Fig. \ref{fig3} a). This is a flow finite size effect, as opposed to a flow finite time effect: for us, the relevant adimensional Weissenberg parameter $f_0/\Delta \gamma$ (with $\Delta \gamma$ the gas damping, see below) remains always much larger than 1 \cite{Ekinci1}.

The molecular damping is linear in pressure $P$, and well documented in 
room temperature experiments \cite{Ekinci2}. However, at cryogenic temperatures a deviation has been reported: the measured gas damping is smaller than expected \cite{Rasul,Martial}. This has been attributed to a {\it rarefaction phenomenon} occurring in the boundary Knudsen layer of the gas \cite{Rasul}. Physically, when $\lambda_{mfp}/g$ grows above 1 the atoms in the gas thermalise not only through collisions among themselves, but also with the nearby wall (at distance $g$ below the device). In the limit $\lambda_{mfp}/g \gg 1$, the scattering from the boundary surface dominates the damping mechanism. The difference between room temperature and low temperature results is presumably linked to physical properties of adsorbed atoms (forming from 1 - 3 disordered layers on the surfaces) \cite{Rasul,Noury}.
Therefore, with our devices which are fully suspended (i.e. $g \rightarrow \infty$), this effect should not exist and we should recover the linear in $P$ law. This is indeed demonstrated in Fig. \ref{fig3} a, which confirms the conclusions reached in Refs. \cite{Rasul,Martial}.

However, with beams of width larger than $200 \ \mathrm{nm}$ ($500$ and $1000 \ \mathrm{nm}$ here), we observe again a (rather small) decrease of the damping below the molecular law (see Fig. \ref{fig3} b). In order to demonstrate this effect, we plot the gas damping $\Delta \gamma$ (i.e. with mechanical damping contribution subtracted from measured peak width) normalised to the molecular law $a \, P$.
The mechanical linewidth was carefully measured prior to gas introduction using the same magnetic field. Note that for the smallest $w$ device ($100 \ \mathrm{nm}$, magenta color in Fig. \ref{fig3}), an addendum contribution of about $+20 \ \mathrm{Hz}$ (on top of the initial $250 \ \mathrm{Hz}$) had to be taken into account: an extra damping ({\it above} the standard molecular expectation) was visible at the smallest pressures.
The following phenomenological equation is fit to the data ($a, h, l, b$ positive real numbers):
\begin{equation}
    \frac{\Delta \gamma}{a \, P} = \frac{1}{1 + h\cdot P + \frac{l}{1 + b \cdot P}} , \label{fitequa}
\end{equation}
which is chosen to preserve reasonable asymptotic behaviors: it falls off at high and low  pressures (as $1/P$ and $P$ respectively), while remaining finite (and reaching $\approx 1$ around a Torr). 
With this expression, the $P \rightarrow 0$ limiting value is $0< 1/(1+l) < 1$. This choice is different from the one performed in Ref. \cite{Rasul} using a similar phenomenology, and discussed explicitly below.

\begin{figure}[h]
\centering
\includegraphics[width=\linewidth]{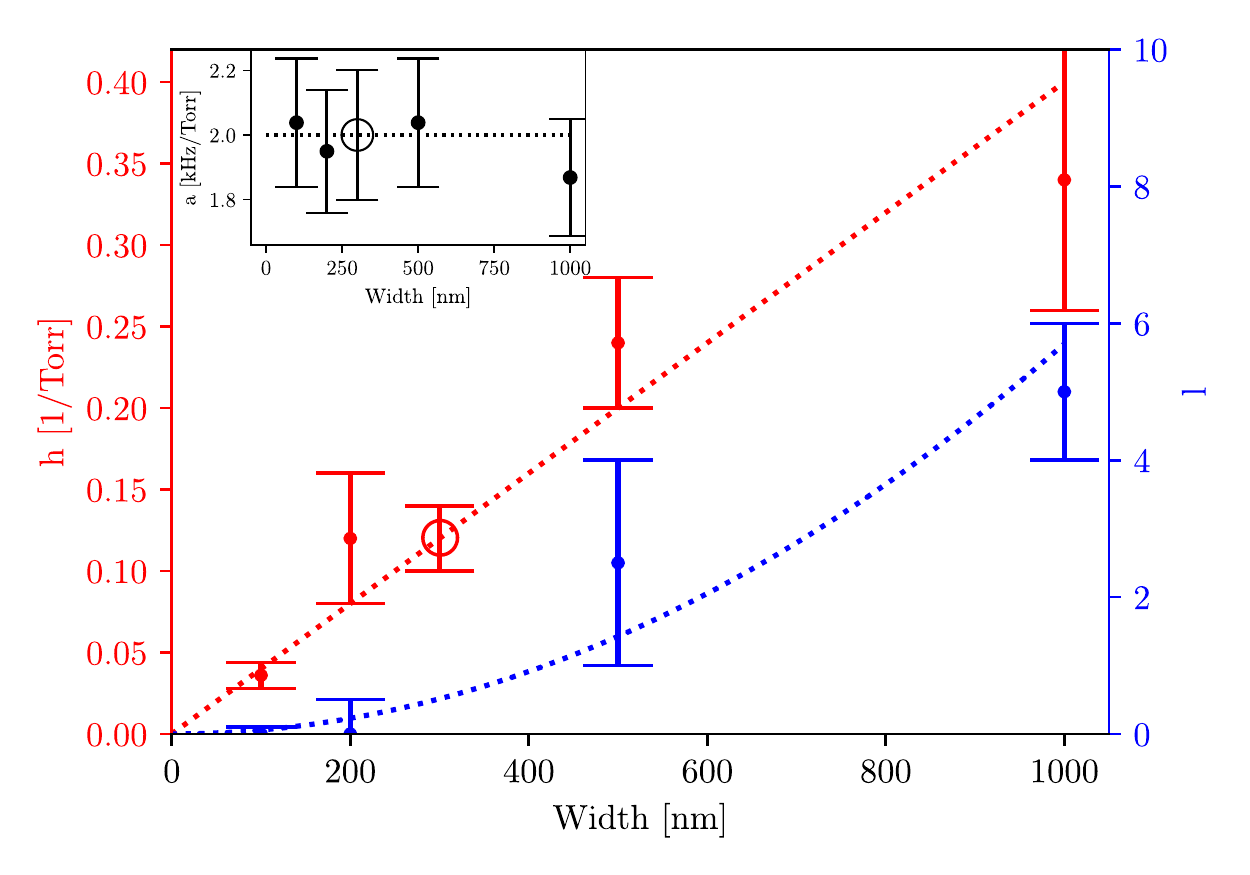}
\caption{{\bf Fit parameters from Eq. (\ref{fitequa})}
\newline
Main graph: parameter $h$ capturing the molecular-to-laminar transition at high pressures (red points, left axis), and parameter $l$ used to fit the low pressure deviation (blue points, right axis). For this fit, the value of $b$ was kept fixed at $b = 300 \ \mathrm{Torr}^{-1}$ (see text). The $x$ axis is the width $w$ of the device. 
Inset: parameter $a$ as a function of width $w$, which defines the amplitude of the gas damping around $1 \ \mathrm{Torr}$ where both low and high pressure deviations are negligible (i.e. $\Delta \gamma \approx a \, P$). The empty circles correspond to the device of Ref. \cite{Rasul}. Dashed lines are simple guides to the eyes (see discussion in text).
}
\label{fig4}
\end{figure}

The fit parameters from Eq. (\ref{fitequa}) are plotted in Fig. \ref{fig4} as a function of width $w$. Since the thickness $e$ is the same for all devices (about $130 \ \mathrm{nm}$), these graphs can also be interpreted as a dependence on aspect ratio $w/e$ (ranging from about 1 to 10).

Although Eq. (\ref{fitequa}) contains 4 fit parameters, we should point out that they are constrained in very different pressure ranges. The $a$ coefficient gives the overall amplitude of the damping around $1 \ \mathrm{Torr}$, where all deviations from the ideal molecular $\propto P$ law are essentially negligible. It is shown in Fig. \ref{fig4} inset, and seems to be almost constant within our error bars, with possibly a slight decrease for the broadest beam.

The $h$ parameter captures the molecular-to-laminar crossover at high pressures. Eq. (\ref{fitequa}) creates a simple interpolation from $\Delta \gamma \approx a \, P$ to $\Delta \gamma \approx a/h$ as $P$ increases, which enables to link (at lowest order) the molecular law with the proper high-pressure Navier-Stokes solution \cite{Sader} (represented in Fig. \ref{fig3} b with dashed lines). As a result, $h$ seems to be essentially linear with the width $w$, a feature reminiscent of the discussion presented at the beginning of this section about the finite size nature of the cross-over. This empirical procedure is rather different from published works \cite{Ekinci1,Ekinci2}. Although it is very practical, the fits performed here are not very accurate because of the reduced high pressure range explored in our work (see error bars in Fig. \ref{fig4}). 
The remaining parameters from Eq. (\ref{fitequa}) describe the small device-dependent decrease seen at low pressures, Fig. \ref{fig3} b. 
The analytic expression we use is slightly different from the one used in Ref. \cite{Rasul} which was focused on explaining a $\Delta \gamma \propto P^2$ dependence in the Knudsen boundary layer; here instead, we propose a $\Delta \gamma \propto (A+ B \, P) \, P$ law when $P \ll 1 \ \mathrm{Torr}$. This has been done on pure empirical grounds: we found that fits were better with $A \neq 0$. 
However, the drawback of our expression is that $l$ and $b$ coefficients are quite correlated within the fitting routine. We therefore decided to fix $b$ in order to decrease the number of variables to adjust (note that this could not be performed in a reasonable manner the other way round, by fixing $l$).

The values of $l$ (for $b = 300 \ \mathrm{Torr}^{-1}$) are presented in Fig. \ref{fig4} as a function of $w$. Error bars are quite large from the fit (note that $b$ can also be chosen $\pm 100\ \mathrm{Torr}^{-1}$
producing a similar outcome). However, the {\it tendency} is the relevant result here: for small devices (i.e. $w < 300 \ \mathrm{nm}$ typically)
there is no fitable correction, while for larger ones it is clearly growing.
We interpret this phenomenon as being the signature of the Knudsen boundary layer rarefaction occurring {\it onto the device itself}. 
In the limit of an infinitely wide probe, one should indeed recover the result of Ref. \cite{Rasul} when deeply in the boundary layer, namely when $\lambda_{mfp}/g \gg 1$ (and the $g$ parameter becomes irrelevant). 
From the fit, we effectively obtain $A=1/(1+l)$ decreasing with $w$ and reaching a value as low as 0.15 for the $1 \ \mu \mathrm{m}$ beam, which is consistent with the data of Ref. \cite{Rasul} presented in Fig. \ref{fig3} a.
However, the quality of the data does not allow to decide whether 
or not the asymptotic ``infinite wall" solution ($w \rightarrow \infty$) shall be strictly $A=0$.
The meaning of $l$ plotted in Fig. \ref{fig4} is that there is a $^4$He related lengthscale to which $w$ should be compared to, in order 
to define if the probe is non-invasive or not (in the sense of a local change in the spatial distribution of the gas properties).
This cannot be $\lambda_{mfp}$, which already defined the cross-over to the molecular regime (see discussion at the beginning of the Section). 
What shall it then be? The mean-free-path of particles in the adsorbed layers? In Fig. \ref{fig3} c we present the normalised added mass $\Delta m/m$ computed from the frequency shifts $\Delta f_r$ observed when gas is introduced:
\begin{equation}
\frac{\Delta f_r}{f_r} =- \frac{1}{2}\frac{\Delta m}{ m} .
\end{equation}
It is therefore clear that in the whole pressure range of the experiment, at least one solid layer of $^4$He covers the NEMS devices. 


As a conclusion to this section, from the guide to the eyes in Fig. \ref{fig4} we infer that $l$ cannot be much larger than 0.75 for a $300 \ \mathrm{nm}$ wide beam. This implies about 15 \% max. damping reduction at $10^{-2}$ Torr, which is within the error bars of the measurements and demonstrates that this effect does not impact the conclusions of Ref.  \cite{Rasul}. 
Note that non-invasivity can also be understood in terms of the absence of disturbance of the {\it momentum distribution} of gas properties. This is excluded here, since experiments are conducted at low enough nano-beam velocities, such that no nonlinear damping effects can be noticed \cite{Rasul, Martial}.

\section{Conclusion}
\label{sec13}

In this work we present the realisation of new probes for low and ultra-low temperature investigations of quantum fluids, which are fully suspended over a hollow window in the silicon chip, and which have cross dimensions as small as $50 \ \mathrm{nm}$.
The mechanical properties have been carefully characterised at $4 \ \mathrm{K}$, demonstrating very good quality, consistent with expectation for such NEMS devices.
We report on the straightforward implementation of multiplexing of nano-beams signals, with independent response of up to 5 devices connected in series.

A benchmark experiment has been realised in rarefied gaseous $^{4}\mathrm{He}$.
We defined the transition from molecular to laminar flow, as a function of the aspect ratio of the probe, in agreement with the  theoretical predictions for a rectangular cylinder immersed in a viscous fluid \cite{Sader}.

The experiment confirms the findings of Ref. \cite{Rasul}: in the absence of a nearby wall, for a probe device having essentially the same cross section, the decrease in gas damping attributed to the boundary layer rarefaction effect is not seen. The molecular regime presents a friction force which remains linear in pressure in the range we studied.

However, a small decrease in damping is observed for the largest devices of width $500 \ \mathrm{nm} $ and $1000 \ \mathrm{nm} $; but not for smaller ones. We interpret it as a boundary layer effect occurring on the device itself: the beam surface is large enough to be ``the wall", leading to a rarefaction effect which mitigates the friction. 
In this sense, for this specific experiment, only probes with cross dimensions below $300 \ \mathrm{nm}$ shall be considered as truly non invasive.

From the results reported in this article, we believe that these new types of NEMS devices will enable new experiments to be conducted in the field of quantum fluids, potentially unravelling some of the key puzzles unsolved so far.
\backmatter


\bmhead{Acknowledgments}
We acknowledge the use of the N\'eel Nanofab facility, and fruitful discussions with Rasul Gazizulin, Benjamin Pigeau and Jean-Philippe Poizat.
The authors acknowledge the support from ERC StG grant UNIGLASS No. 714692, and ERC CoG grant ULT-NEMS No. 647917.
The research leading to these results has received funding from the European Union's Horizon 2020 Research and Innovation Programme, under grant agreement No. 824109, the European Microkelvin Platform (EMP).

\bibliography{sn-bibliography}

\end{document}